\documentclass[12pt,letterpaper]{article}
\usepackage{epsfig,rotating,setspace,latexsym,amsmath,epsf,amssymb,amsfonts,bm,theorem,cite,enumerate,longtable,accents,float,physics,algorithm,graphicx,epsf,authblk,epstopdf,url,xcolor,soul,multirow,bbm,mathrsfs,subcaption,mathtools,comment,tree-dvips,hyperref,dsfont}
\usepackage[center]{qtree}
\usepackage[linguistics]{forest}
\usepackage[noend]{algpseudocode}

\newtheorem{theorem}{Theorem}

\newtheorem{remark}{Remark}

\newtheorem{example}{Example}

\allowdisplaybreaks

\date{}

\begin{document}
\title{HetDAPAC: Leveraging Attribute Heterogeneity in Distributed Attribute-Based Private Access Control\footnote{A part of this work was presented at IEEE ISIT, 2024.}}

\author{Shreya Meel \qquad Sennur Ulukus\\
\normalsize Department of Electrical and Computer Engineering\\
\normalsize University of Maryland, College Park, MD 20742 \\
\normalsize {\it smeel@umd.edu} \qquad {\it ulukus@umd.edu}}

\maketitle

\vspace*{-1.2cm}

\begin{abstract}
Verifying user attributes to provide fine-grained access control to databases is fundamental to an attribute-based authentication system. In particular, the verifying authority lets the user access the records it is eligibile for, only after it verifies the possession of these attributes at the user. This can be realized in two ways: Either a single (central) authority verifies all the attributes, or multiple independent authorities verify the attributes distributedly. In the central setup, the central authority verifies the user attributes, and the user downloads only the single record it can access. While this is communication efficient, it reveals all user attributes to the central authority. A distributed setup with multiple authorities prevents this privacy breach by allowing each authority to verify and learn a single user attribute. Motivated by this, recently Jafarpisheh~et~al.~studied the distributed setup by introducing an information-theoretic formulation of the problem, known as \emph{distributed attribute-based private access control} (DAPAC). With $N$ non-colluding authorities (servers), $N$ attributes and $K$ possible values for each attribute, the $(N,K)$ DAPAC system lets each server learn only the single attribute value that it verifies, and remains oblivious to the remaining $N-1$ attributes. Further, the user retrieves its designated record, without learning any information about the remaining database records. The goal is to maximize the \emph{rate}, i.e., the ratio of the size of desired message to the size of total download from all servers. However, not all attributes are sensitive in practice, and the privacy constraints in DAPAC can be too restrictive, negatively affecting the rate. To leverage the heterogeneous privacy requirements of user attributes, we propose heterogeneous (Het)DAPAC, a framework which off-loads the verification of $N-D$ out of $N$ attributes to a central server, and retains the DAPAC architecture with $D$ dedicated servers for the $D$ sensitive attributes. First, we provide an achievable HetDAPAC scheme, which improves the rate from $\frac{1}{2K}$ in Jafarpisheh~et~al.~to $\frac{1}{K+1}$, i.e., almost doubles the rate while sacrificing the privacy of a few possibly non-sensitive attributes. Unlike DAPAC, our scheme entails an imbalance in the amount of downloads, across the servers. We quantify this asymmetry, and propose a second achievable scheme with a more balanced cost per server, achieving the rate of $\frac{D+1}{2KD}$.
\end{abstract}

\section{Introduction}
Attribute-based encryption (ABE) \cite{watersABE, penuelas2023revocation} is an authentication mechanism wherein a central authority verifies the \emph{attributes} of a user, in order to grant the user access to their data. ABE is widely adopted as a cryptographic protocol in cloud-based storage systems \cite{saeedPH3R, wangABE} to support data protection and authenticated access to multiple enterprises. Here, a user, identified by its set of attributes, wishes to access its designated record stored in the database. It can retrieve only the record it is qualified for, by getting its attributes verified by the authority. ABE can be implemented in a centralized or distributed manner. In a centralized system, a single authority is entrusted to verify all attributes. This is harmful for the privacy of user attributes, since the combination of all attributes completely exposes the user to the authority. The non-centralized system overcomes this by distributing the task of verifying user attributes among multiple, independent authorities. This is known as multi-authority attribute-based encryption \cite{chaseMABE}. Such systems are studied using cryptographic primitives such as bilinear mapping, encryption algorithms and hashing \cite{jungprivateMABE, wangABE, saeedPH3R, han2014ppdcp}. All these works are based on computational hardness and do not provide information-theoretic privacy guarantees.
 
Reference \cite{aliDAPAC} is the first paper to formulate an information-theoretic framework for the multi-authority attribute-based encryption system in the form of distributed attribute-based private access control (DAPAC) system. A DAPAC system comprises multiple non-communicating authorities (servers) and a single user, where the user requesting its data record from the authorities, is uniquely identified by its attribute vector. Its goal is to prevent each server from learning any information about the attributes that it does not verify. While the security and privacy constraints are similar to those of private information retrieval (PIR) \cite{jafarPIR, tian2019capacity, ulukusPIRLC} and its symmetric variant SPIR \cite{jafarSPIR}, there are two key differences: First, the attribute value committed to each server decides the records (messages) that the user can access, including its designated message. This is unlike the canonical PIR problem where all messages are equally accessible, and all messages are eligible to be retrieved by the user. The accessibility pattern results in a non-replicated storage structure, with only the message replicated in all databases being the requested one. Further, the interplay between the accessible database messages and the required message index in DAPAC, renders the model different from the variants of PIR, where the storage across databases is non-replicated, and is described by a graph (e.g., \cite{banawanNONREPIR, ravivNONREPIR, jafar4starPIR, asymp_gxstpir, meel_multi_pir, meel2025symmetric}). Second, unlike PIR where the message index is completely hidden from every server, a single entry of the user's attribute vector is observed by each server. The constraint is to preserve the privacy of all the remaining attributes.

\textbf{Our contributions:} In this work, we explore an $(N,D,K)$ heterogeneous DAPAC (HetDAPAC) system where a subset of $D$ attributes are verified by $D$ distributed authorities (one attribute per authority), while a single central authority verifies the remaining $(N-D)$ attributes and relays their values to the $D$ distributed authorities. For example, some generic user attributes, such as, country of residence, age, zipcode can be non-sensitive information that can be collectively shared with multiple authorities. One of the organizations can serve as the central authority by verifying these attributes and relaying them to the rest, who verify only one sensitive attribute each of the user. This way, few potentially non-sensitive (or generic) attributes of the user are revealed to all the authorities, still preserving the privacy of the sensitive user attributes. This system is \emph{heterogeneous} both in terms of the verification responsibilities of the authorities, as well as the privacy requirements of the attributes. This natural \emph{heterogeneity} in user attributes poses a relaxed privacy constraint, enabling us to improve the communication efficiency compared to that of \cite{aliDAPAC}. 

We present a HetDAPAC scheme which increases the achievable \emph{rate} to $\frac{1}{K+1}$, almost doubling the DAPAC rate of $\frac{1}{2K}$. In this scheme, the user requests for a single linear combination of its accessible messages from each dedicated server, and $KD$ linear combinations from the central server. The message is retrieved in the form of $D$ sub-packets (chunks of message symbols). Each sub-packet is recovered through interference cancellation between the central and one dedicated server. We introduce \emph{load ratio} as a metric to quantify the asymmetry in download costs between the two types of servers. To alleviate the heavier download cost of the central server, compared to each dedicated server, we propose a time-sharing approach. This is derived by noticing that the $(D,K)$ DAPAC scheme of \cite{aliDAPAC} is also an achievable  $(N,D,K)$ HetDAPAC scheme, where the central server only verifies the generic attributes and relays them to the remaining servers, while the user follows the scheme in \cite{aliDAPAC} with the $D$ dedicated servers only. This allows us to establish HetDAPAC rates for a given load ratio, with the tradeoff of diminishing rate with more balanced download costs.

Next, we improve the tradeoff achieved through time-sharing, by proposing a second $(N,D,K)$ HetDAPAC scheme, which achieves the rate $\frac{D+1}{2KD}$. This scheme involves an intricate query design approach, where the user requests for multiple linear combinations from all servers. In particular, the designated message is retrieved in the form of $\binom{D+1}{2}$ sub-packets through interference cancellation performed between the downloads of the dedicated servers, as well as between those of the central and each dedicated server. The resulting download cost from the central server is $\frac{D}{D-1}$ times that of each dedicated server. This equips us with the characterization of an improved tradeoff between rate and load ratio, that is achievable through time-sharing our two proposed schemes and the scheme in \cite{aliDAPAC}. 

The rest of the paper is organized as follows. Section \ref{sec:bkground and sysmod} briefly overviews the DAPAC system and describes the new HetDAPAC system model. We state our results in Section \ref{sec:results}. In Section \ref{ach1} and Section \ref{ach2}, we present the two achievable schemes. Finally, in Section \ref{sec:conclude}, we conclude the paper with some directions for future work.

\section{Background and System Model}\label{sec:bkground and sysmod}
In this section, we present a brief overview of the DAPAC system of \cite{aliDAPAC}, and describe our HetDAPAC system model. 

\subsection{DAPAC System} 
In an $(N,K)$ DAPAC system, we use the attribute vector $\bm{v}^*=(v_1^*,v_2^*,\ldots,v_N^*)$ to represent the user's $N$ independent attributes. Each attribute $v_n^*$ takes one out of $K$ possible values from a corresponding finite alphabet set $\mathcal{V}_n$, with $|\mathcal{V}_n|=K$, $n\in [N]$, where $[N]:=\{1,\ldots,N\}$. The servers store the message set $\mathcal{W}$ of $K^N$ messages in a replicated manner, each message $W_{\bm{v}}\in \mathcal{W}$ corresponding to a unique attribute vector $\bm{v}$. In addition, the servers also share a set $\mathcal{S}$ of common randomness symbols, independent of ${\bm{v}}^*$, $\mathcal{W}$ which is unavailable to the user. Each $W_{\bm{v}}$ is an independent message of $L$ symbols from a finite field $\mathbb{F}_q$. Further, the symbols in $\mathcal{S}$ are uniformly and independently picked from $\mathbb{F}_q$. 

The user commits $v_n^*$ to server $n$ and on verification, is given access to a subset of messages of $\mathcal{W}$. Based on the user's query, each server releases the requested combinations of accessible messages to the user, secured by a common randomness symbol. From the queries and answers, the user exactly recovers $W_{\bm{v}^*}$, while gaining no information on $\mathcal{W}\setminus W_{\bm{v}^*}$. The queries are designed such that server $n$ obtains no information on the user's attribute vector beyond $v^{*}_n$. The goal is to design a scheme that achieves this with a minimum cost of downloading answers per symbol of $W_{\bm{v}^*}$. Equivalently, the rate of an $(N,K)$ DAPAC system, defined as the ratio of message length to the download cost, should be maximized. Jafarpisheh~et~al.~\cite{aliDAPAC} show that the rate $\frac{1}{2K}$ is achievable. We illustrate their scheme's idea with an example.

\begin{example} \label{example1}
    Consider a $(3,2)$ DAPAC system with $N=3$ attributes ${\bm{v}}=(v_1,v_2,v_3)$ with $v_1 \in \mathcal{V}_1=\{a,b\}$, $v_2 \in \mathcal{V}_2=\{1,2\}$ and $v_3 \in \mathcal{V}_3=\{x,y\}$ and each message having $L=3$ symbols from a finite field $\mathbb{F}_q$. Let ${\bm{v}^*}=(v_1^*,v_2^*,v_3^*)=(a,2,y)$ be the particular attribute of the user.\footnote{For instance, $v_1$ could be the gender attribute with ``$a$'' being female and ``$b$'' being male, $v_2$ could be the education level with ``$1$'' being BS and ``$2$'' being PhD, and $v_3$ could be the major with ``$x$'' being CS and ``$y$'' being ECE. In this case, the person with features $(a,2,y)$ is a female with a PhD in ECE.} Thus, the designated message for the user is $W_{a2y}$. The user sends $v_1^*=a$ to server 1, $v_2^*=2$ to server 2, and $v_3^*=y$ to server 3. Once verified, the messages accessible to the user at each server are,
    \begin{align}
      \text{Server 1:} & \qquad \{W_{a1x}, W_{a1y}, W_{a2x}, W_{a2y}\}, \nonumber \\
      \text{Server 2:} & \qquad \{W_{a2x}, W_{a2y}, W_{b2x}, W_{b2y}\}, \nonumber \\
      \text{Server 3:} & \qquad \{W_{a1y}, W_{a2y}, W_{b1y}, W_{b2y}\}.
    \end{align}
    This is shown in Fig.~\ref{fig:example 1 servers}. We observe that this produces a non-uniform replication pattern, where $W_{a2y}$ (the desired message) is replicated on all three servers, $W_{a2x}$, $W_{b2y}$ and $W_{a1y}$ are replicated on two servers each, and $W_{a1x}, W_{b2x}$ and $W_{b1y}$ are not replicated at all.

    The user privately chooses a random permutation $\pi_{\bm{v}}:[3]\to[3]$ to rearrange the symbols of  each accessible message $W_{\bm{v}}$, with $[w_{\bm{v}}(1) \, w_{\bm{v}}(2) \, w_{\bm{v}}(3)]$ denoting the permuted symbols. Then, the user generates $12$ random vectors $h_{nm} ,\ n\in[3], \ m\in [4]$ uniformly from $\mathbb{F}_q^2$, $9$ of which are independent, and  $3$ of which satisfy, $h_{21}=h_{12}+e_2$, $h_{31} = h_{14}+e_2$ and $h_{34} = h_{24}+ e_1$, with $e_j$ denoting the $2\times 1$ column vector with zeros in all but the $j$th row, $j=1,2$. As a query to server $n$, user sends the vectors $h_{nm}, \ m\in [4]$ and the permuted indices of symbols to be linearly combined with $h_{nm}$. All operations are performed in $\mathbb{F}_q$. The set of answers downloaded is shown in Table~\ref{tabex12}. Here, a ``$;$'' between message sub-packets denotes that they are stacked column-wise, and $s_j \in \mathbb{F}_q, \ j\in[9]$ are distinct parts of the shared common randomness $\mathcal{S}$.
    
    \begin{figure}[t]
        \centering
        \includegraphics[width=0.6\textwidth]{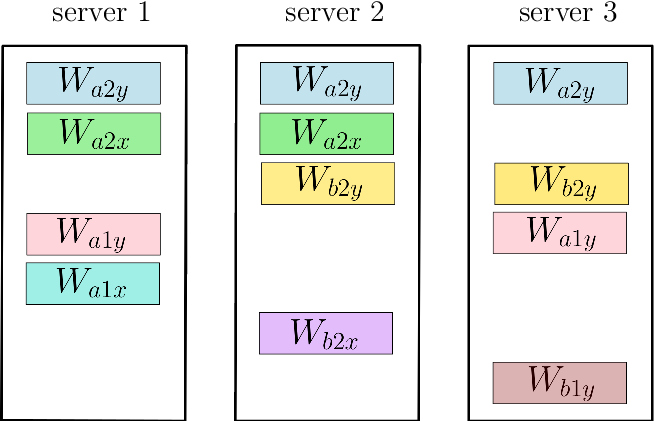}
        \caption{Accessible messages in $(3,2)$ DAPAC system of Example~\ref{example1}, exhibiting a non-uniform replication pattern.}
        \label{fig:example 1 servers}
    \end{figure}
    
    \begin{table}[htbp]
    \centering
    \begin{tabular}{|c|c|}
    \hline
    server & answer\\
    \hline
    \multirow{4}{*}{server 1} & $h_{11} ^T [w_{a1x}(1) ; w_{a1y}(1)] +s_1$ \\ 
    & $h_{12} ^T [w_{a2x}(1) ; w_{a2y}(1)]+s_2$ \\ 
    & $h_{13} ^T [w_{a1x}(2) ; w_{a2x}(2)] +s_3$ \\ 
    & $h_{14} ^T [w_{a1y}(2) ; w_{a2y}(2)] +s_4$ \\
    \hline
    
    \multirow{4}{*}{server 2} & $(h_{12} +e_2) ^T [w_{a2x}(1) ; w_{a2y}(1)] +s_2$ \\ 
    & $h_{22} ^T[w_{a2x}(3) ; w_{b2x}(1)]+s_5$ \\ 
    & $h_{23} ^T [w_{b2x}(2) ; w_{b2y}(1)] +s_6$ \\ 
    & $h_{24} ^T [w_{a2y}(3) ; w_{b2y}(2)] +s_7$ \\
    \hline
    
    \multirow{4}{*}{server 3} & $(h_{14} + e_2)^T [w_{a1y}(2) ; w_{a2y}(2)] +s_4$ \\ 
    & $h_{32} ^T [w_{a1y}(1) ; w_{b1y}(1)]+s_8$ \\ 
    & $h_{33} ^T [w_{b1y}(2) ; w_{b2y}(2)] +s_9$ \\ 
    & $(h_{24} +e_1) ^T [w_{a2y}(3) ; w_{b2y}(2)] +s_7$ \\
       \hline
    \end{tabular}
    \caption{Answers downloaded from the servers.}
    \label{tabex12}
    \end{table}
    
    By subtracting the second answer of server 1 from the first answer of server 2, the user obtains the symbol $w_{a2y}(1)$ of $W_{a2y}$. The two remaining symbols are similarly obtained by pairing answers of servers $1$, $3$ and servers $2$, $3$. Thus, to recover $3$ message symbols, the user downloads $12$ symbols, yielding the $(3,2)$ DAPAC rate of $\frac{3}{12}=\frac{1}{4}$ which is $\frac{1}{2K}$, as $K=2$. Also, the secrecy of message symbols other than $W_{a2y}$ is maintained, by virtue of one-time padding \cite{shannon_otp} in the downloaded answers.
\end{example}

In general, the achievable $(N,K)$ DAPAC scheme splits each message into $\binom{N}{2}$ equal chunks (sub-packets). From each server, the user requests for $K(N-1)$ linear combination of distinct message sub-packets, each of which involving messages corresponding to two common attributes, e.g., $v_n^*$ and $v_m, m\neq n$ from server $n$. Using the $KN(N-1)$ downloads, the user recovers the $\binom{N}{2}$ sub-packets of the desired message. The achievable rate is $\frac{N(N-1)/2}{KN(N-1)}=\frac{1}{2K}$.

\subsection{HetDAPAC System}
Similar to the $(N,K)$ DAPAC system, the user is identified by $N$ attributes and wishes to access its designated/desired record (message) stored in multiple servers. Different from \cite{aliDAPAC}, while we allow a subset of $D$ out of $N$ attributes to be verified by dedicated distributed servers (one attribute verified per server), we allow the remaining $(N-D)$ attributes to be verified by the $(D+1)$st, i.e., the central, server. The $(N-D)$ attributes, after verification, are relayed by the central server to the $D$ dedicated servers. Thus, in our $(N,D,K)$ HetDAPAC system, $(N-D)$ attributes are publicly known to all $(D+1)$ servers in the system, while each dedicated server learns and verifies only one sensitive attribute. If $N-D=0$, by omitting the central server, our $(N,D,K)$ HetDAPAC system reduces to the $(N, K)$ DAPAC system. 

Under this HetDAPAC setup, the user wishes to download its designated message $W_{\bm{v}^*}$, corresponding to its attribute vector $\bm{v}^*$. This is accomplished in two phases:

\subsubsection{Verification Phase} The user sends its attribute $v^*_{n},n\in [D]$, to server $n$, and the remaining attributes, $v^*_{D+1}, \ldots, v^*_N$, to server $(D+1)$ for verification. Server $(D+1)$ verifies the $(N-D)$ attributes and sends them to the $D$ servers, while remaining oblivious to the attribute values $v^*_1, \ldots, v^*_D$. 

\begin{figure}[t]
    \centering
    \includegraphics[width=0.5\linewidth]{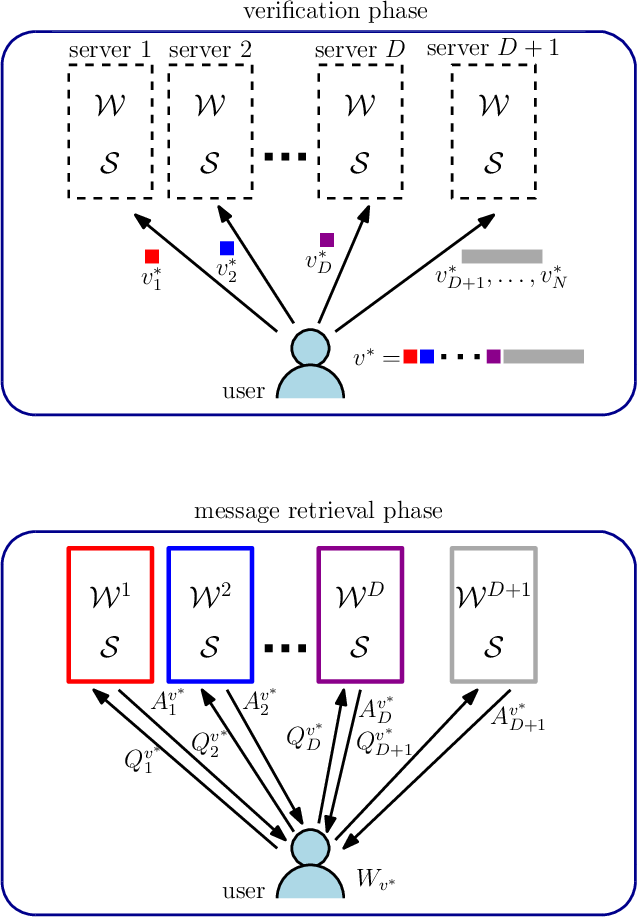}
    \caption{Two phases of an $(N,D,K)$ HetDAPAC system.}
    \label{model}
\end{figure}

\subsubsection{Message Retrieval Phase} Note that, in server $(D+1)$, there are $K^D$ accessible messages, corresponding to attribute vectors whose attribute values $v_{D+1},\ldots,v_N$ are fixed to $v^*_{D+1},\ldots,v^*_N$; and in server $n\in[D]$, there are $K^{D-1}$ messages corresponding to attribute vectors for which $v_n = v^*_n$ and $v_{D+1},\ldots,v_N=v^*_{D+1},\ldots,v^*_N$. We call the message repository in server $n$, as database $n$, and represent it by $\mathcal{W}^n$. 

Our system model is shown in Fig.~\ref{model}. From the user's perspective, the access control policy imposes a non-colluding \emph{non-replicated} storage across databases, with the following properties: i) only the designated message $W_{\bm{v}^*}$ is replicated in all databases, ii) any pair of dedicated databases share $K^{D-2}$ common messages, iii) the content of database-$n$ is a subset of the content of database-$(D+1)$, for any $n\in [D]$. While properties i) and ii) are applicable to an $(N,K)$ DAPAC system with $D=N$, iii) additionally occurs in our proposed $(N,D,K)$ HetDAPAC framework.

Given the accessible storage, the user sends queries\footnote{The problem formulation of \cite{aliDAPAC} considers the user to have access to an independent permutation random variable $\mathcal{P}$, in addition to the queries in the access control constraints. We, on the other hand, absorb $\mathcal{P}$ into the queries $Q^{\bm{v^*}}_{[1:D+1]}$.} $Q_n^{\bm{v}^*}$  to server $n\in [D+1]$. The queries are generated without the knowledge of the database content, thus,
\begin{align}
     I(Q^{\bm{v}^*}_1,\ldots,Q^{\bm{v}^*}_{D+1}; \mathcal{W})=0.
\end{align}
The queries sent and the answers generated should be such that the correctness, attribute privacy and database secrecy constraints are satisfied. We assume that the servers respond to the user with answers if and only if the user queries for messages in its accessible database. The correctness and database secrecy together form the access control constraints. For correctness, the designated message $W_{\bm{v}}^{*}$ should be recoverable using all the queries and answers, i.e.,
\begin{align}\label{eq:reliability}
\text{[Correctness]} \quad   H(W_{\bm{v}^{*}}|Q^{\bm{v}^*}_{[1:D+1]}, A^{\bm{v}^*}_{[1:D+1]})=0.
\end{align}
Next, we define the attribute privacy constraints. To each dedicated server $n\in [D]$, no information on the $(D-1)$ remaining attributes of the user should be leaked from the query $Q^{\bm{v}^*}_n$ to server $n$. Similarly, to the $(D+1)$st server, the user's query $Q^{\bm{v}^*}_{D+1}$ should reveal no information on any other attribute, i.e.,
\begin{align}\label{eq:user_priv1}
\text{[Attribute Privacy]} \quad &I(Q^{\bm{v}^*}_n; v^{*}_{[1:D]}\backslash v^{*}_n|\mathcal{S}, \bm{v}^{*}_{[D+1:N]},v^{*}_n, \mathcal{W}) = 0.\\
\label{eq:user_priv2}
& I(Q^{\bm{v}^*}_{D+1}; v^{*}_{[1:D]}|\mathcal{S},v^{*}_{[D+1:N]},  \mathcal{W}) =0. 
\end{align}
Here, \eqref{eq:user_priv1} and \eqref{eq:user_priv2} allow the knowledge of the last $(N-D)$ attributes to all servers. In response to the queries, each server $n\in [D+1]$ sends the answer $A^{\bm{v}^*}_n$ to the user, which is a function of $Q^{\bm{v}^*}_n$, $\mathcal{W}^n$ and $\mathcal{S}$, 
\begin{align}
    H(A_n^{\bm{v}^*}|Q^{\bm{v}^*}_n, \mathcal{W}^n, \mathcal{S}) = 0.
\end{align} 
Note that $\mathcal{S}$ is independent of $\mathcal{W}$ and $v^*_1,\ldots,v^*_D$; it is shared based on the common knowledge of $v^*_{D+1},\ldots,v^*_{N}$ among the servers. Finally, to ensure database secrecy, the user should not learn anything about the remaining messages  beyond the message it is supposed to learn, denoted by the set $\mathcal{W}_{\bm{\overline{v}}^*}=\mathcal{W}\setminus W_{\bm{v^*}}$, i.e.,
 \begin{align}\label{eq:db_secrecy}
 \text{[Database Secrecy]} \quad    I(\mathcal{W}_{\bm{\overline{v}}^*}; Q^{\bm{v}^*}_{[1:D+1]}, A^{\bm{v}^*}_{[1:D+1]}) =0.
 \end{align}

A feasible $(N,D,K)$ HetDAPAC scheme is one that satisfies \eqref{eq:reliability}, \eqref{eq:user_priv1}, \eqref{eq:user_priv2} and \eqref{eq:db_secrecy}. The \emph{rate} $R$ of an $(N,D,K)$ HetDAPAC system is the ratio of the message length $L$ to the total download cost from all servers, and the goal is to maximize this. The capacity of $(N,D,K)$ HetDAPAC system is the maximum achievable rate over all feasible schemes. In general, the capacity of HetDAPAC is greater than the capacity of DAPAC system. This is because of the more stringent privacy requirement of DAPAC, with equal treatment of all $N$ attributes. In this perspective, the privacy requirement of HetDAPAC is relaxed, resulting in a higher capacity, compared to that of DAPAC for the same $N$ and $K$.

We assume symmetry in download costs across all dedicated servers. To characterize the asymmetry in the download costs between each dedicated and the central server, we define load ratio, $\ell$ as the following ratio
\begin{align}
    \ell=\frac{\text{number of downloaded symbols from any server $n\in [D]$}}{\text{number of downloaded symbols from server $D+1$}},
\end{align}
for a given $(N,D,K)$ HetDAPAC scheme. Depending on the communication settings, it might be desired to have a flexible load ratio in the message retrieval phase. Therefore, another goal of the HetDAPAC problem is to characterize the optimal rate-load ratio trade-off, i.e., maximum achievable rate for a given load ratio. 

\section{Results}\label{sec:results}
In this section, we state the main results of our paper.
\begin{theorem}\label{thm1}
    For any $(N,D,K)$ HetDAPAC system, the following rate and load-ratio pair is achievable,
    \begin{align}
        R=\frac{1}{K+1}, \qquad\ell = \frac{1}{KD}, \label{thm1-eqn}
    \end{align}
with a minimum required common randomness size $H(\mathcal{S})\geq KL$.
\end{theorem}

The proofs of Theorem~\ref{thm1} and the subsequent remark follow from the first proposed scheme, presented in Section~\ref{ach1}.

\begin{remark}\label{rmk1}
    Similar to the scheme in \cite{aliDAPAC}, the rate of our scheme $R$ in \eqref{thm1-eqn} is independent of the number of attributes $N$ and the number of databases $(D+1)$. This is because, both download cost and sub-packetization grow linearly with $D$, i.e., with the number of private attributes. Hence, their ratio is constant in $D$ and decreases only with the alphabet size $K$. 
\end{remark}

\begin{remark}\label{rem:dapac_feasible}
    If server $(D+1)$ participates only in the verification phase, and not in the message retrieval phase, then the scheme in \cite{aliDAPAC} is a valid achievable scheme for our problem with $N=D$. The resulting rate is $R=\frac{1}{2K}$, at a load ratio $\ell=\infty$, with a minimum common randomness size of $H(\mathcal{S})\geq K^2L$.    
\end{remark}

\begin{remark}\label{rmk3}
    By time-sharing between our first scheme and the $(D,K)$ DAPAC scheme, any rate-load ratio pair $(R(\lambda),\ell(\lambda) )$ between $\left(\frac{1}{K+1},\frac{1}{KD}\right)$ and $\left(\frac{1}{2K},\infty\right)$ is achievable, for $\lambda\in [0,1]$. The following equation  specifies the trade-off between rate and load ratio,
     \begin{align}\label{eq:tradeoff}
        R(\lambda) = \frac{1}{K+\frac{\ell(\lambda)K^2D+K}{\ell(\lambda)KD+2K-1}}.  
    \end{align}
\end{remark}
The proof of Remark~\ref{rmk3} is straightforward and is presented in Section~\ref{proof_time_share}. 

\begin{theorem}\label{thm2}
    For an $(N,D,K)$ HetDAPAC system with $D\geq 3$, the following rate and load-ratio pair is achievable,
    \begin{align}
        R=\frac{D+1}{2KD}, \qquad\ell = \frac{D-1}{D},
    \end{align}
with a minimum required common randomness size $H(\mathcal{S})\geq \frac{D-1}{D+1} K^2L$.
\end{theorem}

\begin{remark}\label{rem4}
The rate in Theorem \ref{thm2} is greater than the rate achieved through time-sharing (Remark \ref{rmk3}) for the same load ratio $\frac{D-1}{D}$. Moreover, the improvement in rate over the time-shared scheme decreases with $K$ and $D$.
\end{remark}
The proofs of Theorem~\ref{thm2} and Remark~\ref{rem4} are a consequence of the second achievable scheme.

\begin{figure}[t]
    \centering
    \includegraphics[width=0.6\linewidth]{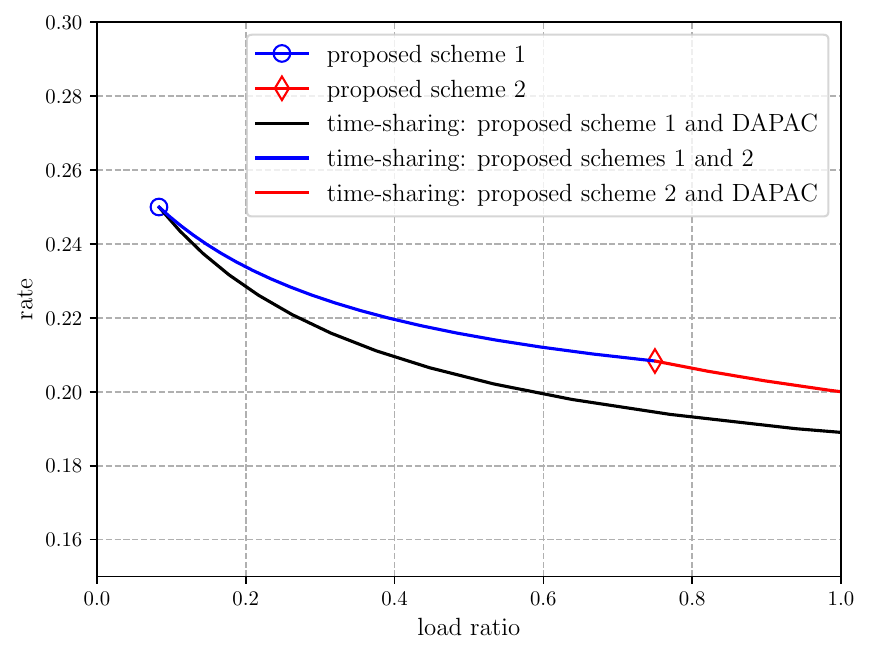}
    \caption{Achievable rate-load ratio pair for $(N,4,3)$ HetDAPAC system.}
    \label{fig:results}
\end{figure}

\begin{remark}\label{rem5}
By time-sharing between our first and second proposed schemes, and the scheme in \cite{aliDAPAC}, we obtain an improved rate-load ratio trade-off, compared to those obtained in Remark \ref{rmk3}, as illustrated in Fig.~\ref{fig:results}.
\end{remark}
The proof of Remark~\ref{rem5} follows similarly to that of Remark \ref{rmk3}, by time-sharing between our first scheme and the second scheme for $\ell\in \left(\frac{1}{KD},\frac{D-1}{D}\right)$ and between the second scheme and that of \cite{aliDAPAC} for $\ell\in \left(\frac{D-1}{D},\infty\right)$.

\section{Proof of Theorem \ref{thm1}: Proposed Scheme 1}\label{ach1}
In this section, we present the achievable scheme that proves Theorem \ref{thm1}. Before proceeding with the general scheme description, we illustrate it with an example of $(3,2,2)$ HetDAPAC system, with the same attributes as in Example~\ref{example1}.

\begin{example}\label{example 2}
    Let each message consist of $L=2$ symbols in $\mathbb{F}_q$ and $\bm{v}^*=(a,2,y)$. After the verification phase, the messages accessible to the user from the databases are,
    \begin{align}
        \text{Server 1:} & \qquad \{W_{a1y}, W_{a2y}\}, \nonumber \\
        \text{Server 2:} & \qquad \{W_{a2y}, W_{b2y}\}, \nonumber \\
        \text{Server 3:} & \qquad \{W_{a1y}, W_{a2y}, W_{b1y}, W_{b2y}\}.
    \end{align}
    This is shown in Fig.~\ref{fig: example 2 servers}. The user privately chooses a random permutation $\pi_{\bm{v}}:[2]\to[2]$ to rearrange the symbols of  each accessible message $W_{\bm{v}}$, with $[w_{\bm{v}}(1) \, w_{\bm{v}}(2)]$ denoting the permuted symbols. Next, the user concatenates the sub-packets of messages to be linearly combined from each server into message groups. From the central database (server 3), the user requests for four message groups $w_{nk}$ where $ n\in [2], \, k \in [2]$. From server 1 and server 2, the message groups $w_1:=w_{11}$ and $w_2:=w_{22}$, are requested. Further, the user chooses $4$ independent random vectors $h_{11}, h_{12}, h_{21}, h_{22}$ uniformly from $\mathbb{F}_q^{2\times 1}$ as linear combination coefficients. Using these, the user queries the databases and receives the responses as given in Table~\ref{tabex22}. Here, $s_{nk}$, $n,k\in [2]$ are chunks of $\mathcal{S}$ shared among the servers.

    \begin{figure}[t]
       \centering
       \includegraphics[width=0.6\textwidth]{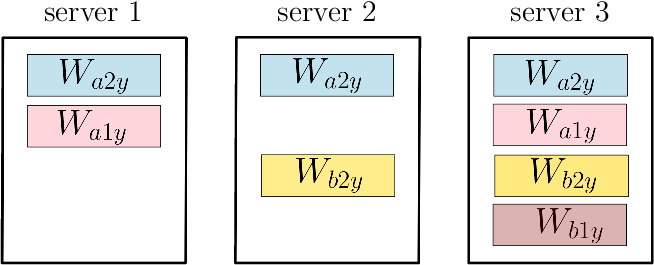}
       \caption{Accessible messages in $(3,2,2)$ HetDAPAC system of Example~\ref{example 2}, exibiting a non-uniform replication pattern.}
       \label{fig: example 2 servers}
    \end{figure}
    
    \begin{table}[htbp]
    \centering
    \begin{tabular}{|c|c|c|}    
    \hline
    server & message group & answer\\
    \hline
    server 1 & $ w_{1} = [w_{a1y}(1) ; w_{a2y}(1)]$ & $( {h_{11}} + {e_2})^T w_{1}  + s_{11} $\\
    \hline
    server 2 & $ w_2 = [w_{a2y}(2) ; w_{b2y}(2)]$ & $( {h_{22}}+ {e_1}) ^T  w_{2}+ s_{22} $ \\
    \hline
    & $w_{11}=[w_{a1y}(1) ; w_{a2y}(1)]$ &  $ {h_{11}}^Tw_{11}  + s_{11} $\\
    server 3 & $w_{12}=[w_{b1y}(1) ; w_{b2y}(1)]$ & ${h_{12}}^T w_{12} + s_{12} $\\
    & $ w_{21}=[w_{a1y}(2) ; w_{b1y}(2)]$ &  $ {h_{21}}^T w_{21}  + s_{21} $\\
    & $ w_{22}=[w_{a2y}(2) ; w_{b2y}(2)]$ &  $ {h_{22}}^T w_{22}  + s_{22} $\\
    \hline
    \end{tabular}
    \vspace*{0.2cm}
    \caption{Queries and answers for $ {\bm{v}}^*=(a, 2, y)$.}
    \label{tabex22}
    \end{table}
    
    The user can subtract the first and fourth answers of server 3 from the answers of server 1 and server 2 to recover the $2$ symbols of $W_{a2y}$, yielding the rate $\frac{2}{1+1+4}=\frac{1}{3}$, i.e., $\frac{1}{K+1}$. The corresponding load ratio is $\frac{1}{4}$ and $4$ symbols of common randomness are required. Since servers 1 and 2 are each requested for a single symbol of user-accessible messages, while server 3 is requested for two symbols each of user-accessible messages, attribute privacy is preserved. Further, no information about the remaining message symbols is revealed to the user since they are protected by parts of the server common randomness.
\end{example}

\subsection{Proposed Scheme 1}
Now, we describe the achievable scheme resulting in the rate $\frac{1}{K+1}$ which proves Theorem~\ref{thm1}. Each message in $\mathcal{W}$ is split into $D$ equal sub-packets of $L/D$ symbols each. Thus, $W_{\bm{v}}= [W_{\bm{v}}(1) \ W_{\bm{v}}(2) \ \ldots \ W_{\bm{v}}(D)]$, where each sub-packet $W_{\bm{v}}(i)$ is a row vector of $L/D$ symbols from $\mathbb{F}_q$. For each accessible message corresponding to ${\bm{v}}$, the user privately chooses an independent random permutation $\pi_{{\bm{v}}}:[D]\to [D]$ from  the set of all permutations $\Pi$, $|\Pi|=D!$  to rearrange the sub-packets. Let $w_{\bm{v}}(i)= W_{\bm{v}}(\pi_{\bm{v}}(i))$ and $[w_{\bm{v}}(1) \ w_{\bm{v}}(2) \ \ldots \ w_{\bm{v}}(D)]$ denote the permuted sub-packets. 

\subsubsection{Query Construction} 
We represent the attribute alphabet set $\mathcal{V}_n$ as an ordered set, where $\mathcal{V}_n(k)$ is the $k$th value in the set. This order is known to the user and the servers. The query design involves assigning the message groups whose linear combination is desired by the user and designing the corresponding linearly combining vectors. We begin with the queries sent to server $(D+1)$. For each $n\in[D]$, $k\in [K]$, consider the set of messages,  
\begin{align}\label{eq:define_U}
    \mathcal{U}(n,k)=\left\{W_{\bm{v}} \mid v_n=\mathcal{V}_n(k), \ v_{[D+1:N]}=v^*_{[D+1:N]}\right\}.
\end{align}
Clearly, $|\mathcal{U}(n,k)|=K^{D-1}$. For each message in $\mathcal{U}(n,k)$, the user chooses a new sub-packet index and vertically concatenates these message sub-packets to obtain the message group $w_{nk}$. Each $w_{nk}$ is composed of $K^{D-1}$ sub-packets from accessible messages, and is represented as a $K^{D-1}\times (L/D)$ matrix in $\mathbb{F}_q$. Thus, there are $KD$ message groups for server $(D+1)$. Further, the user generates $KD$ random vectors $h_{nk}, n\in [D]$, $k\in [K]$, of dimension $K^{D-1}\times 1$, whose elements are picked uniformly and independently from $\mathbb{F}_q$. The user sends the tuple of the $K$ message group and linear combining vectors as query to the server.

Now, we describe the query design for each dedicated server $n\in [D]$. Let the user attribute $v^*_n=\mathcal{V}_n(k_n)$. Consequently, we have that $\mathcal{W}^n = \mathcal{U}(n,k_n)$, which is given as:
\begin{align}
    \mathcal{W}^n &=\left\{W_{\bm{v}}\mid v_n=v_n^*, \ v_{[D+1:N]}=v^*_{[D+1:N]}\right\}.
\end{align}
For $n\in[D]$, we assign a single message group $w_n$ defined as,
\begin{align}
   w_{n}=w_{nk_n}.
\end{align}
Let $l_n\in [K^{D-1}]$ be the position of the $w_{\bm{v^*}}(n)$ in $w_{nk_n}$. Then, the corresponding linear-combining vector is,
\begin{align}\label{eq:define h_n}
   h_n=h_{nk_n} + e_{l_n},
\end{align}
where $e_{l_n}$ is the unit column vector of length $K^{D-1}$ with $1$ at the $l_n$th coordinate. Thus, a single pair of message group and linear combining vector is sent as query to server $n\in [D]$.

Therefore, the query tuple $Q_n^{{\bm{v}}^*}$ for server $n$ is,
\begin{align}
    Q_n^{{\bm{v}}^*} =
    \begin{cases}
        (h_n, w_n), & n\in [D],\\
        \{(h_{mk},w_{mk}), m\in [D], k\in [K]\}, & n=D+1.
    \end{cases}
\end{align}

\subsubsection{Answer Generation}  
The servers split $\mathcal{S}$ into chunks of $L/D$ symbols and assign one sub-packet to each $\mathcal{U}(n,k)$, $n\in[D]$, $k\in [K]$, through an injective mapping, and label this as $s_{nk}$. This is done prior to the verification of sensitive attributes $v^*_{[1:D]}$. To generate answers, each server multiplies the vectors in the query tuple with the queried message group and adds a chunk of $\mathcal{S}$ to it, as follows,
\begin{align}
        A_n^{\bm{v}^*} =  
    \begin{cases}
        h_n^T w_n +s_{nk_n}, &   n\in [D],\\
        \{h_{mk}^T w_{mk} + s_{mk}, m \in [D], k \in [K]\}, &   n=D + 1.
    \end{cases}   
\end{align}
The subscript of $s_{nk}, n\in [D], k\in [K]$ corresponds to the message group parameters $w_{nk}$. For each $\mathcal{U}(n,k)$, $\frac{L}{D}$ distinct sub-packets of $\mathcal{S}$ are used, resulting in $H(\mathcal{S})\geq KL$.

Next, we show that the scheme satisfies reliability, attribute privacy and database secrecy.

\paragraph{Correctness} To obtain $w_{\bm{v}^*}(n)$, the user subtracts the answer corresponding to the query $(h_{nk_n}, w_{nk_n})$ sent to server $(D+1)$ from the answer of the dedicated server $n$. That is, for each $n\in [D]$,
\begin{align}
    w_{\bm{v}^*}(n) &= (h_n^Tw_n + s_{nk_n})-(h_{nk_n}^T w_{nk_n}+s_{nk_n})\\
    &= (h_{nk_n}+e_{l_n})^Tw_{nk_n}-h_{nk_n}^Tw_{nk_n}\label{eq:extract message}\\
    &=e_{l_n}^Tw_{nk_n},
\end{align}
where \eqref{eq:extract message} follows from \eqref{eq:define h_n}, and since $w_{\bm{v}^*}(n)$ is the $l_n$th row of the message group $w_{nk_n}$. This way, the $D$ sub-packets of $W_{\bm{v}^*}$ are recovered and rearranged with $\pi_{\bm{v}^*}^{-1}$. The resulting rate is,
\begin{align}
    R&=\frac{D}{KD+D}=\frac{1}{K+1}.
\end{align}
The user downloads a single sub-packet from each dedicated server, and $KD$ sub-packets from server $(D+1)$, resulting in the load ratio of $\ell=\frac{1}{KD}$.

\paragraph{Attribute Privacy} Each dedicated server uses the coefficients $h_n$ to linearly combine the components of $w_n$, which is composed of one sub-packet each, of all messages in $\mathcal{W}^n$. Hence, from the perspective of the dedicated server $n$, the user attributes $v^*_{[1:D]\setminus \{n\}}$ are completely obfuscated. Similarly, from the central server, the user downloads $KD$ linear combinations of the accessible messages, independent of $v^*_{[1:D]}$. The order of the sub-packet indices are hidden in the requested linear combinations, through independent permutations $\pi_{\bm{v}}$ from all servers. Finally, since the servers are non-colluding, the corresponding linear combining vectors appear to be uniformly random vectors from $\mathbb{F}_q$. As a result, the query tuples respect the attribute privacy constraint in \eqref{eq:user_priv1} and \eqref{eq:user_priv2}.

\paragraph{Database Secrecy} Each answer from server $n$ is a linear combination of message sub-packets added to a randomness sub-packet, unavailable to the user. If $W_{\bm{v}^*}$ does not appear in a linear combination, then it is not useful for decoding. Hence, no information on $\mathcal{W}^n \setminus W_{\bm{v}^*}$ for $n\in [D+1]$ is revealed to the user. This ensures that the access control constraints are also maintained. 

We illustrate our scheme using another example.
\begin{example}\label{ex:2_2}
    Consider an $(4,3,2)$ HetDAPAC system. Let $\mathcal{V}_1=\{a,b\}, \mathcal{V}_2 = \{1,2\}, \mathcal{V}_3 = \{u,v\}$ and $\mathcal{V}_4=\{x,y\}$. Let $\bm{v}^* = (a,2,u,y)$ and $L=D=3$ for each message. The accessible messages for the user in the message retrieval phase are,
    \begin{align}
    \text{Server 1:} & \qquad \{W_{a1uy}, W_{a2uy}, W_{a1vy}, W_{a2vy}\}, \nonumber \\
    \text{Server 2:} & \qquad \{W_{a2uy}, W_{b2uy}, W_{a2vy}, W_{b2vy}\}, \nonumber \\
    \text{Server 3:} & \qquad \{W_{a1uy}, W_{b1uy}, W_{a2uy}, W_{b2uy}\}, \nonumber\\
    \text{Server 4:} & \qquad \{W_{a1uy}, W_{b1uy}, W_{a2uy}, W_{b2uy}, W_{a1vy}, W_{b1vy}, W_{a2vy}, W_{b2vy}\}.
    \end{align}
    The resulting message subsets $\mathcal{U}(n,k)$, $n\in [3], k\in [2]$, are,
    \begin{align}\label{eq:msg_subsets_ex}
        \mathcal{U}(1,1)&=\{W_{au1y},W_{a2uy},W_{a1vy},W_{a2vy}\},\notag\\
        \mathcal{U}(1,2)&=\{W_{b1uy},W_{b2uy},W_{b1vy},W_{b2vy}\},\notag\\
        \mathcal{U}(2,1)&=\{W_{a1uy},W_{b1uy},W_{a1vy},W_{b1vy}\},\notag\\
        \mathcal{U}(2,2)&=\{W_{a2uy},W_{b2uy},W_{a2vy},W_{b2vy}\},\notag\\
        \mathcal{U}(3,1)&=\{W_{a1uy},W_{b1uy},W_{a2uy},W_{b2uy}\},\notag\\
        \mathcal{U}(3,2)&=\{W_{a1vy},W_{b1vy},W_{a2vy},W_{b2vy}\}.   
    \end{align}
    Accordingly, the servers allocate the common randomness symbols $s_{n,k}\in \mathcal{S}$ for each $\mathcal{U}(n,k)$. This determines the queries and answers for the central server as shown in Table \ref{tab:ex2_3}. Notice that $k_1=1,k_2=2,k_3=1$. Consequently, the message group for server $n$ corresponds to the message subset $\mathcal{U}(n,k_n)$ as given in \eqref{eq:msg_subsets_ex}. The sub-packets of $W_{a2uy}$ appear in the $2$nd, $1$st and $3$rd rows of message groups of servers 1, 2 and 3, respectively. This determines the query tuple and the corresponding  answers of the dedicated servers.
    
    \begin{table}[htbp]
    \centering
    \begin{tabular}{|c|c|c|}    
    \hline
    server & message group & answer\\
    \hline
    server 1 & $ w_{1} = [w_{a1uy}(1) ; w_{a2uy}(1);w_{a1vy}(1) ; w_{a2vy}(1)]$ & $( {h_{11}} + {e_2})^T w_{1}  + s_{11} $\\
    \hline
    server 2 & $ w_2 = [w_{a2uy}(2) ; w_{b2uy}(1); w_{a2vy}(2) ; w_{b2vy}(1)]$ & $( {h_{22}}+ {e_1}) ^T  w_{2}+ s_{22} $ \\
    \hline
    server 3 & $w_{3}=[w_{a1uy}(2) ; w_{b1uy}(1); w_{a2uy}(3); w_{b2uy}(2)]$ &  $ (h_{31}+e_3)^T w_{3}  + s_{31} $\\
    \hline
    \multirow{6}{*}{server 4} & $ w_{11} = w_1$ & ${h_{11}}^T w_{11}  + s_{11} $\\
    & $w_{12}=[w_{b1uy}(2) ; w_{b2uy}(1); w_{b1vy}(1); w_{b2vy}(2)]$ & ${h_{12}}^T w_{12} + s_{12} $\\
    & $ w_{21}=[w_{a1uy}(2) ; w_{b1uy}(1) ; w_{a1vy}(2); w_{b1vy}(1)]$ &  $ {h_{21}}^T w_{21}  + s_{21} $\\
    & $ w_{22}= w_2$ &  $ {h_{22}}^T w_{22}  + s_{22} $\\
    & $w_{31}=w_3$ & ${h_{31}}^T w_{31} + s_{31} $\\
    & $ w_{32}=[w_{a1vy}(2) ; w_{b1vy}(2) ; w_{a2vy}(3) ; w_{b2vy}(3)]$ &  $ {h_{32}}^T w_{32}  + s_{32} $\\
    \hline
    \end{tabular}
    \caption{Queries and answers for $ {\bm{v}}^*=(a, 2, u, y)$.}
    \label{tab:ex2_3}
    \end{table}
    
    The user downloads $1$ symbol each from the first 3 servers and $6$ symbols from server 4, i.e., $9$ symbols total to recover $3$ symbols of $W_{a2uy}$, yielding a rate $\frac{1}{3}$, i.e., $\frac{1}{K+1}$, load ratio $\frac{1}{6}$ and using $6=2\cdot3$ symbols of common randomness.
\end{example}

\subsection{Proof of Remark \ref{rmk3}: Time-Sharing}\label{proof_time_share}
  Let us fix $\lambda\in(0,1)$. Given $\lambda$, we apply the scheme in \cite{aliDAPAC} to the first $\lambda L$ symbols and the scheme in Section~\ref{ach1} to the remaining $(1-\lambda)L$ symbols of each message to obtain a valid $(N,D,K)$ HetDAPAC scheme. Then, the download cost from each dedicated server is
  \begin{align}\label{eq:dedicated_download}
     \lambda\cdot K(D-1)\frac{L}{\binom{D}{2}}+ (1-\lambda)\cdot\frac{L}{D} = \left((2K-1)\lambda+1\right)\cdot\frac{L}{D},
  \end{align}
while that from server $(D+1)$ is 
  \begin{align}\label{eq:central_download}
    (1-\lambda)\cdot KD\frac{L}{D}=(1-\lambda)KL.
  \end{align}
Thus, the achievable rate as a function of $\lambda$ is,
\begin{align}\label{eq:rate_ts}
    R(\lambda) &= \frac{L}{L\cdot\left( (2K-1)\lambda+1+K(1-\lambda)\right)}\\
    &=  \frac{1}{K(1+\lambda)+(1-\lambda)}.
\end{align}
The two extreme points of \eqref{eq:rate_ts} correspond to $R(1)=\frac{1}{K+1}$ and $R(0)=\frac{1}{2K}$. 
The amount of common randomness $\mathcal{S}$ to be shared at the servers for a given $\lambda$ is
\begin{align}\label{eq:randmoness_ts}
    H(\mathcal{S},\lambda) = KL(\lambda(K-1)+1).
\end{align}
Using \eqref{eq:dedicated_download} and \eqref{eq:central_download}, the load ratio $\ell(\lambda)$,
\begin{align}
    \ell(\lambda) = \frac{2K\lambda + 1 - \lambda}{KD(1-\lambda)}=\frac{1}{KD} + \frac{2\lambda}{D(1-\lambda)}. \label{load_ratio_lambda}
\end{align}
Solving for $\lambda$ in \eqref{load_ratio_lambda} and
substituting it in \eqref{eq:rate_ts} yields
\begin{align}
    R(\lambda) &= \frac{1}{K+\frac{\ell(\lambda)K^2D+K}{\ell(\lambda)KD+2K-1}}.
\end{align} 
This time-sharing results in the achievable rate-load ratio pairs as depicted by the black curve in Fig.~\ref{fig:results}.

\section{Proof of Theorem \ref{thm2}: Proposed Scheme 2}\label{ach2}
In this section, we propose a scheme that attains the rate $\frac{D+1}{2KD}$, which is greater than the time-shared scheme at the load ratio of $\frac{D-1}{D}$ for $D\geq 3$. Our scheme solves the imbalance in download costs with a load ratio approaching $1$ as $D$ increases, while improving the rate. This scheme utilizes the message overlap between every pair of the $D+1$ databases more efficiently than time-sharing, i.e., the $K^{D-2}$ messages between dedicated servers, and $K^{D-1}$ between each dedicated and the central server. Consequently, if $D=2$, there is no overlap in messages between any pair of dedicated servers. Hence, our scheme does not provide a gain beyond time-sharing. That is why we require $D\geq 3$. We motivate our scheme with the following example of a $(4,3,2)$ HetDAPAC system.

\begin{example}\label{ex:3}
    Consider the same settings as in Example \ref{ex:2_2} with the set of accessible messages after the verification phase given by \eqref{eq:msg_subsets_ex}. Each message consists of $L=6$ symbols from the finite field $\mathbb{F}_q$. Let $W_{\bm{v}}= [W_v(1) \ldots W_v(6)]$. The servers share $12$ common randomness symbols $\mathcal{S}=\{s_1, \ldots, s_{12}\}$, uniformly picked at random from $\mathbb{F}_q$. 
   
    The user applies private permutations to the indices of the messages it can access from the databases, to hide the order of retrieving them. For instance, let $W_{\bm{v}}$ be denoted by $[w_v(1) \ldots w_v(6)]$ after permuting the indices. The user requests for specific linear combinations of the message groups as given in Table \ref{tabex3} from each server. For the dedicated servers, each message group constitutes the symbols corresponding to attribute vectors with two attributes in common, while for the central server, each message group constitutes those with one attribute in common, apart from $y$. For each message group $w_{i,j}$, the user generates uniform random vectors $h_{i,j}$ independently from $\mathbb{F}_q^8$. Note that, the same $h_{i,j}$ is assigned to certain pairs of message groups, i.e., $h_{2,1}=h_{1,2}$, $h_{3,1}=h_{1,3}$ and $h_{3,4}=h_{2,3}$. For server 4, three message groups are formed by concatenating the existent message groups and three new message groups, $w_{4,1},w_{4,2}, w_{4,3}$ are created as shown in Table \ref{tabex3}. The corresponding answers from all servers are generated by linearly combining the received query vector with the requested message group and summing it to a suitable randomness symbol. Notably, server $4$ adds a sum of two randomness symbol to each of its linear combinations.
    
    \begin{table}[htbp]
    \centering
    \begin{tabular}{|c|c|c|}
    \hline
    server & message group & answer\\
    \hline
    \multirow{4}{*}{server 1}& $w_{1,1}= [w_{a1uy}(1) ; w_{a1vy}(1)]$ &  ${h_{1,1}}^Tw_{1,1}  + s_{1} $\\
    & $w_{1,2}= [w_{a2uy}(1) ; w_{a2vy}(1)]$ &  ${{h_{1,2}}^T w_{1,2} + s_{2} }$\\
    & $ w_{1,3} = [w_{a1uy}(2) ; w_{a2uy}(2)]$ &  ${{h_{1,3}}^T w_{1,3} + s_{3} }$\\
    & $ w_{1,4}= [w_{a1vy}(2) ; w_{a2vy}(2)]$ &  ${h_{1,4}}^T w_{1,4} + s_{4} $\\
    \hline
    
    \multirow{4}{*}{server 2} & $ w_{2,1}= [w_{a2uy}(4) ; w_{a2vy}(1)]$ &  $ {{h_{1,2}}^T w_{2,1}  + s_{2} }$\\
     & $ w_{2,2}= [w_{b2uy}(1) ; w_{b2vy}(1)]$ &  ${h_{2,2}}^T w_{2,2}+ s_{5} $\\
     & $ w_{2,3} = [w_{a2uy}(3) ; w_{b2uy}(2)]$ &  $ {{h_{2,3}}^T w_{2,3}  + s_{6} }$\\
     & $ w_{2,4} =  [w_{a2vy}(2) ; w_{b2vy}(2)]$ &  $ {h_{2,4}}^T w_{2,4} + s_{7} $\\
    \hline
    
     \multirow{4}{*}{server 3} & $ w_{3,1} =  [w_{a1uy}(2) ; w_{a2uy}(5)]$ &  $ {{h_{1,3}}^T w_{3,1} + s_{3} }$\\
    & $ w_{3,2} = [w_{b1uy}(1) ; w_{b2uy}(3)]$ &  ${h_{3,2}}^T w_{3,2} + s_{9} $\\
    & $ w_{3,3}= [w_{a1uy}(3) ; w_{b1uy}(2)] $ &  $ {h_{3,3}}^T w_{3,3}  + s_{8} $\\
    
    & $ w_{3,4} =  [w_{a2uy}(6) ; w_{b2uy}(2)] $ &  $ { {h_{2,3}}^T w_{3,4} + s_{6} }$\\
    \hline
     \multirow{6}{*}{server 4} & $w_{4,1} = [w_{1,1}; w_{1,2}]$ &  {$h_{1,1}^T w_{1,1} + (h_{1,2}+e_1)^T w_{1,2} +(s_1 +s_2)$}\\
       &$ w_{4,2} = [w_{b1uy}(3) ; w_{b1vy}(1); w_{b2uy}(4) ; w_{b2vy}(3)]$ & $h_{4,2}^T w_{42}+ (s_5 + s_{10})$\\
      &$w_{4,3} = [w_{a1uy}(4) ; w_{b1uy}(4) ; w_{a1vy}(3); w_{b1vy}(2)]$ & $  h_{4,3}^T w_{4,2} + (s_8 +s_{11})$\\
       &$w_{4,4} = [w_{2,3};w_{2,4}]$ & $(h_{2,3}+e_1)^T w_{2,3} + h_{2,4}^T w_{2,4} +(s_6 +s_7)$\\

       &$w_{4,5} = [w_{3,1};w_{3,2}]$ &  $(h_{1,3}+e_2)^T w_{3,1} + h_{3,2}^T w_{3,2} +(s_3 +s_9)$\\
       &$w_{4,6}= [w_{a1vy}(4) ; w_{a2vy}(3); w_{b1vy}(3) ; w_{b2vy}(4)]$ & $h_{4,6}^T w_{4,6} + (s_4 + s_{12})$\\
     \hline
    \end{tabular}
    \caption{Queries and answers for $ {\bm{v}}^*=(a, 2,u, y)$.}
    \label{tabex3}
    \end{table}

    The user decodes all $6$ symbols of $W_{a2uy}$ as follows. To decode $w_{a1uy}(1)$, the user computes the sum of the first two answers of server 1 and subtracts it from the first answer of server 4. Again, by subtracting the second answer of server 1 from the first answer of server 2, the user obtains $h_{12}^T(w_{a2uy}(4)-w_{a2uy}(1))$ from which it recovers $w_{a2uy}(4)$. Next, by computing the sum of the third and fourth answers of server 2 and subtracting it from the fourth answer of server 4, the user decodes $w_{a2uy}(3)$. By subtracting the third answer of server 2 from the last answer of server 3, the user obtains  $h_{2,3}^T(w_{a2uy}(6)-w_{a2uy}(3))$, from which it recovers $w_{a2uy}(6)$. Through a similar operation with the third answer of server 1, first and second answers of server 3 and fifth answer of server 4, the user recovers $w_{a2uy}(2)$ and $w_{a2uy}(5)$. In total, $18$ symbols are downloaded, out of which $3$ symbols from the central server, and $3$ symbols from each dedicated server are useful for decoding $W_{a2uy}$. This yields the rate $\frac{6}{18}=\frac{1}{3}$, i.e., $\frac{D+1}{2KD}$, while the rate by time-sharing is $\frac{7}{24}$ by \eqref{eq:tradeoff}.
\end{example}

\subsection{Proposed Scheme 2}
Now, we describe the general scheme that results in the rate $R=\frac{D+1}{2KD}$ at the load ratio $\ell=\frac{D-1}{D}$. The messages are split into $M=\binom{D+1}{2}$ sub-packets, each comprising $\frac{L}{M}$ symbols. Thus, $W_{\bm{v}}= [W_{\bm{v}}(1) \ W_{\bm{v}}(2) \ \ldots \ W_{\bm{v}}(M)]$ where each sub-packet $W_{\bm{v}}(i)$ is a row vector of $L/M$ symbols from $\mathbb{F}_q$. As in the first scheme, the user assigns private and independent permutation mappings $\pi_{\bm{v}}:[M]\rightarrow[M]$ uniformly at random to the sub-packet indices of each accessible message $W_{\bm{v}}$. We introduce the following notations for the formal description of the message retrieval phase.

Let $\mathcal{P}_D$ denote the collection of all $2$-subsets of $[D]$, hence $|\mathcal{P}_D|=\binom{D}{2}$. We define, for all $\{n,m\} \in \mathcal{P}_D$, assuming $n<m$ without loss of generality, and  $k,k'\in [K]$,
\begin{align}\label{eq:define_pairwise_U}
    \mathcal{U}_{nm}(k,k')=\mathcal{U}_{mn}(k',k)=\mathcal{U}(n,k)\cap \mathcal{U}(m,k'),
\end{align}
where $\mathcal{U}(n,k)$ and $\mathcal{U}(m,k')$ are defined in \eqref{eq:define_U}. Therefore, $\mathcal{U}_{nm}(k,k')= \mathcal{U}_{mn}(k',k)$ denotes the set of $K^{D-2}$ messages whose attribute vectors have fixed $v_n=\mathcal{V}_n(k)$, $v_m=\mathcal{V}_m(k')$ along with $v_{[D+1:N]}=v^*_{[D+1:N]}$. The servers, with the knowledge of $v^*_{[D+1:N]}$, share the common randomness $\mathcal{S}$. For each message set, $\mathcal{U}_{nm}(k,k')$, the servers allocate a chunk of $L/M$ symbols, and label this as $s_{nm}^{(k,k')}=s_{mn}^{(k',k)}$. Therefore, we can rewrite $\mathcal{S}=\{s_{nm}^{(k,k')}, \{n,m\}\in \mathcal{P}_D, k,k'\in [K]\}$, resulting in, 
\begin{align}
    H(\mathcal{S})=\binom{D}{2}K^2\cdot \frac{L}{M}=\frac{D-1}{D+1}K^2L,
\end{align}
to be the minimum required randomness for the feasibility of the scheme. We partition $\mathcal{P}_D$ into two disjoint sets: $\mathcal{C}_{D}$ and $\mathcal{C}'_{D}=\mathcal{P}_D\setminus \mathcal{C}_{D}$. In $\mathcal{C}_{D}$,  we select $D$ elements ($2$-subsets) from $\mathcal{P}_D$, such that, for every $n\in [D]$, there exist exactly two distinct pairs $\{n,m_1\}$ and $\{n,m_2\}\in \mathcal{C}_{D}$ where $m_1,m_2\in [D]$, and $m_1\neq m_2 \neq n$. Equivalently, $\mathcal{C}_D$ is a $1$-$(D,2,2)$ design \cite{stinson_designs}. Hence, $|\mathcal{C}_{D}|=D$ and $|\mathcal{C}'_{D}|=\binom{D}{2}-D=\frac{D^2-3D}{2}$.\footnote{$\mathcal{C}_{D}$ may not be unique for a given $D$. For example, with $D=3$, $\mathcal{C}_{3}=\{\{1,2\},\{1,3\},\{2,3\}\}$ is the only valid set. However, with $D=4$,  $\mathcal{C}_{4}=\{\{1,2\},\{2,3\},\{3,4\},\{1,4\}\}$ and $\mathcal{C}_{4}=\{\{1,3\},\{2,3\},\{2,4\},\{1,4\}\}$ are both valid.} For instance, the \emph{cyclic allotment} $\mathcal{C}_D= \{\{n, n+1\}: n\in [D-1] \}\cup \{D,1\}$ is one such valid partition the user may choose for the query generation.

\subsubsection{Query Construction}
First, we design queries for the dedicated servers. The message group of server $n\in [D]$ constitutes the messages whose attribute vector $\bm{v}$ matches in one attribute other than $v^*_{[D+1:N]}$ and $v^*_n$, i.e., one message group per $\mathcal{U}_{nm}(k_n,k)$, for all $m\in [D]\setminus \{n\}, k\in [K]$. Thus, a total of $K(D-1)$ message groups are assigned per dedicated server. Each message group is composed of $K^{D-2}$ sub-packets, represented as a $K^{D-2}\times L/M$ matrix in $\mathbb{F}_q$. To assign sub-packet indices for message groups, we go sequentially from server $1$ to server $D$. For server 1, we pick a new (permuted) sub-packet index for every message in $\mathcal{U}_{1m}(k_1,k)$ for $m\in [2:D], k\in[K]$ arranged in a fixed order. This completes the message groups for server 1. For servers $n\in [2:D]$, we assign a new sub-packet index to every message in $\mathcal{U}_{nm}(k_n,k)$ whenever $k\neq k_m$. These message groups do not contain any sub-packet of $W_{\bm{v^*}}$. Corresponding to the messages in $\mathcal{U}_{nm}(k_n, k_m)$, except for $W_{\bm{v}^*}$, we assign the same set of sub-packet indices for both servers $n$ and $m$, ensuring that the messages in the message groups corresponding to $\mathcal{U}_{nm}(k_n, k_m)$ and $\mathcal{U}_{mn}(k_m, k_n)$ appear in the same order. For the desired message $W_{\bm{v}^*}$, the sub-packet indices are assigned as follows:
\begin{itemize}
    \item If $\{n,m\} \in \mathcal{C}_{D}$, then \textit{two} distinct sub-packets are assigned. Let them be $w_{\bm{v^*}}(i_{nm,1})$ for the message group in server $n$ and $w_{\bm{v^*}}(i_{nm,2})$ for that in server $m$ with $n<m$. Assume without loss of generality that $i_{nm,1}\in [D]$ and $i_{nm,2}\in [D+1:2D]$. This assigns $2D$ sub-packets of $W_{\bm{v}^*}$ for retrieval.
    
    \item If $\{n,m\} \in \mathcal{C}'_{D}$, then \emph{one} new sub-packet, $w_{\bm{v^*}}(i_{nm})$ is picked and the same index $i_{nm}\in [2D+1:\binom{D+1}{2}]$ is assigned to the message groups in both servers $n$ and $m$. This assigns the remaining $\frac{D^2-3D}{2}$ sub-packets of $W_{\bm{v^*}}$.
\end{itemize} 

Define $\bar{n} = (1,2, \ldots, n-1, n+1, \ldots, D)$ to be the ordered set of integers from $[D]\setminus \{n\}$, arranged in an ascending order. Let $\bar{n}(t)$ denote the $t$th element in $\bar{n}$. We label the $K(D-1)$ message groups of server $n\in [D]$ as,
\begin{align}\label{eq:correspondence of message subset and group}
    w_{n,r+k} := \text{message group corresponding to } \mathcal{U}_{n\bar{n}(t)}(k_n, k) \text{ where } r = (t-1)K,
\end{align}
with $k\in[K]$, $t\in [D-1]$. 

Correspondingly, we pick $K(D-1)$ vectors $h_{n,r+k}
\in \mathbb{F}_q^{K^{D-2}}$ uniformly at random for each server $n\in [D]$, one for each $\mathcal{U}_{n\bar{n}(t)}(k_n,k)$. Let, for servers $n,m\in[D]$  with $n<m$, $m=\bar{n}(i)$ and $n= \bar{m}(j)$ for some $i$ and $j$ in $[D-1]$. Then, the two vectors corresponding to the set $\mathcal{U}_{nm}(k_n,k_m)=\mathcal{U}_{mn}(k_m,k_n)$ in servers $n$ and $m$, should satisfy,
\begin{align}\label{eq:h_vector_dedicated}
    h_{m,(j-1)K+k_n} =
    \begin{cases}
       h_{n,(i-1)K+k_m}, &  \{n,m\}\in \mathcal{C}_{D},\\
       h_{n,(i-1)K+k_m}+e_{l_m}, &\{n,m\} \in \mathcal{C}'_{D},
    \end{cases}
\end{align}
where $e_{l_m}$ is the $K^{D-2}$ length unit column vector with $1$ at the $l_m$th row. Here, we assume that $w_{\bm{v^*}}(i_{nm})$ appears in the $l_m$th row of $w_{n,(i-1)K+k_m}$ and $w_{m,(j-1)K+k_n}$. This completes query assignment for the dedicated servers.

Next, for server $(D+1)$, we create $KD$ message groups, using the subsets $\mathcal{U}(n,k), n\in [D], k\in [K]$, as defined in \eqref{eq:define_U}. We write $\mathcal{U}(n,k)$ as,
\begin{align}\label{eq:msg_set_concat}
    \mathcal{U}(n,k) = \bigcup_{k'=1}^K \mathcal{U}_{nm}(k,k')= \bigcup_{k'=1}^K \mathcal{U}_{mn}(k',k),
\end{align}
where $\{n,m\}$ (or $\{m,n\}$) $\in \mathcal{C}_{D}$. Let us permute the components of each $2$-subset of $\mathcal{C}_{D}$ such that, set of the first and the second entries, in isolation, cover the set $[D]$. This results in a set of ordered pairs $\mathcal{C}^o_{D}$. For instance, if $\mathcal{C}_{3}=\{\{1,2\},\{1,3\},\{2,3\}\}$, $\mathcal{C}^o_{3}=\{(1,2),(2,3),(3,1)\}$. For every $n\in [D]$, such that the tuple $(n,m)\in \mathcal{C}_{D}^o$, we vertically concatenate $K$ message groups corresponding to $\mathcal{U}_{nm}(k_n,k)$ for all $k\in [K]$, in a specific order. Hence, each message group has dimension $K^{D-1}\times \frac{L}{M}$. Let $m=\bar{n}(i)$, then the message group corresponding to $\mathcal{U}(n,k_n)$ is given by,
\begin{align}
    w_{D+1,(n-1)K+k_n} &:= [w_{n,r+1}; w_{n,r+2}; \ldots; w_{n,r+K}], \ \text{ where } r = (i-1)K,
\end{align}
for each $k_n, n\in [D]$. Note that, only the $k_m$th row of $ w_{D+1,(n-1)K+k_n}$ contains a sub-packet of $W_{\bm{v^*}}$, which is $w_{\bm{v}^*}(i_{nm,1})$ if $n<m$, and $w_{\bm{v}^*}(i_{nm,2})$, otherwise. Next,  corresponding to each $\mathcal{U}(n,k)$ where $k\neq k_n$, we form a new message group, by picking a new sub-packet index for every message. Specifically, we concatenate new sub-packets of messages corresponding to the subsets $\mathcal{U}_{n\bar{n}(i)}(k,k')$, $k,k'\in [K], k\neq k_n$, for all $i\in [D-1]$. We label the $(K-1)D$ new message groups for $k\neq k_n$ as, 
\begin{align}\label{eq:label_msg_grp}
    w_{D+1,r+k}: =&\text{ message group corresponding to } \mathcal{U}(n,k), \ \text{  where } r=(n-1)K.
\end{align}

Next, we assign the linearly combining vectors for each $w_{D+1,k+(n-1)K}, k\in [K], n\in [D]$. If $k=k_n$ and  $m=\bar{n}(i)$, where $(n,m)\in \mathcal{C}^o_{D}$, we concatenate the vectors for the dedicated servers as follows,
\begin{align}\label{eq:h_vector for D+1}
       h_{D+1,(n-1)K+k_n} &= [h_{n,r+1}; h_{n,r+2}; \ldots; h_{n,r+k_{m}}+e_{l_m};\ldots;h_{n,r+K}], \ \text{ where $r=(i-1)K$.}
\end{align} 
The $k_m$th row of $h_{D+1,K(n-1)+k_n}$ is $h_{n,k_m+r}$ added to $e_{l_m}$, a unit vector of length $K^{D-2}$. This corresponds to the message group $w_{n,r+k_m}$, whose $l_m$th message symbol is $w_{\bm{v}^*}(i_{nm,1})$ or $w_{\bm{v}^*}(i_{nm,2})$. If $l\neq (n-1)K+k_n, \ \ n\in [D]$, the $(K-1)D$ vectors $h_{D+1,l}$ are selected uniformly at random and independently from $\mathbb{F}_q^{K^{D-1}}$ without any constraints.

Finally, the query tuple sent to the servers are,
\begin{align}
    Q_{n}^{\bm{v}^*}= (h_{n,l},w_{n,l}), \ \text{ with }
    l\in \begin{cases}
        [K(D-1)], & n \in [D],\\
        [KD], & n = D+1.
    \end{cases}
\end{align}
\subsubsection{Answer Generation}
Each dedicated server computes the requested linear combination and adds a common randomness sub-packet to generate $K(D-1)$ answer sub-packets. In particular, the set of answers returned by the dedicated server $n\in [D]$ is,
\begin{align}
A^{\bm{v^*}}_n =
    \{h_{n,l}^Tw_{n,l}+s_{n\bar{n}(i)}^{(k_n,k)}, l= (i-1)K+k, i\in [D-1]\}.
\end{align}

The answer generation for server $(D+1)$ is similar to those of the dedicated servers in terms of taking the product of the requested message group with the respective vector. However, it differs in the common randomness that is added for secrecy. In particular, for the query corresponding to $\mathcal{U}(n,k)$, the $K$ common randomness symbols corresponding to the sets in \eqref{eq:msg_set_concat} are summed, before its addition to the answer. The resulting common randomness $\tilde{s}_{(D+1,l)}$ is given by
\begin{align}\label{eq:randomness_sum}
    \tilde{s}_{D+1,l}=\sum_{k'=1}^K s_{nm}^{(k,k')}, \ \text{ where } l = (n-1)K + k.
\end{align}
where $(n,m)\in \mathcal{C}^o_{D}$.
With this, the answer set returned by DB $(D+1)$ is,
\begin{align}
    A^{\bm{v}^*}_{D+1} = \{h_{D+1,l}^T w_{D+1,l} +\tilde{s}_{D+1,l}, l\in [KD]\}.
\end{align}

Next, we show that the scheme satisfies reliability, attribute privacy and database secrecy.

\paragraph{Correctness}
We prove that the above scheme retrieves all $M$ symbols of $W_{\bm{v}^*}$. Decoding is done in two steps. In the first step, $K$ answers from a dedicated server and a single answer from the central server are evaluated to decode the sub-packets $w_{\bm{v^*}}(i_{nm,j})$, where  $(n,m)\in \mathcal{C}^o_{D}$ and $j=1$ if $n<m$ and $j=2$, otherwise. For every $n\in [D]$, consider the answers of server $n$ corresponding to $\mathcal{U}_{nm}(k_n,k)$ for all $k\in [K]$ where $(n,m)\in \mathcal{C}_{D}^o$. Letting $m=\bar{n}(i)$, these answers are $h_{n,(i-1)K+k}^T w_{n,(i-1)K+k}+s_{nm}^{(k_n,k)}$. Summing the answers over $k\in [K]$ gives
\begin{align}
    \sum_{k=1}^K h_{n,(i-1)K+k}^T & w_{n,(i-1)K+k}+s_{nm}^{(k_n,k)} 
    \nonumber\\
    &= \sum_{k=1}^K h_{n,(i-1)K+k}^T w_{n,(i-1)K+k} + \tilde{s}_{D+1,(n-1)K+k_n},\label{eq:sum_dedicated}
\end{align}
where \eqref{eq:sum_dedicated} follows from \eqref{eq:randomness_sum}. Next, consider the answer of server $D+1$ corresponding to the message set $\mathcal{U}(n,k_n)$,
\begin{align}
   h_{D+1,(n-1)K+k_n}^T & w_{D+1,K(n-1)+k_n} + \tilde{s}_{D+1,(n-1)K+k_n} \notag\\
   &=\sum_{k=1}^Kh_{n,(i-1)K+k}^T w_{n,(i-1)K+k} + e_{l_m}^T w_{n,(i-1)K+k_m}+ \tilde{s}_{D+1,(n-1)K+k_n}, \label{eq:central_answer_expression}
\end{align}
where \eqref{eq:central_answer_expression} is due to \eqref{eq:h_vector for D+1}. Subtracting \eqref{eq:sum_dedicated} from \eqref{eq:central_answer_expression} yields $e_{l_m}^Tw_{n,(i-1)K+k_m}=w_{\bm{v}^*}(i_{nm,j})$, where $j=1$ or $j=2$, depending on whether $n<m$ or $n>m$, respectively.
Evaluating the answers for all $n\in [D]$ and $(n,m)\in \mathcal{C}_{D}^o$ yields $D$ out of $M$ sub-packets.

In the second step, we evaluate the answers from each pair of dedicated servers, $\{n,m\}\in \mathcal{P}_D$. First, we focus on the subset $\mathcal{C}_{D}$. For every $\{n,m\}\in \mathcal{C}_{D}$, from server $n$, assuming $n<m$, we consider the answer corresponding to $\mathcal{U}_{nm}(k_n,k_m)$, where $m=\bar{n}(i)$,
\begin{align}
    h_{n,(i-1)K+k_m}^Tw_{n,(i-1)K+k_m}+s_{nm}^{(k_n,k_m)},
\end{align}
while for server $m$,  we consider the answer corresponding to $\mathcal{U}_{mn}(k_m,k_n)$, where $n=\bar{m}(j)$,
\begin{align}
 h_{m,(j-1)K+k_n}^T&w_{m,(j-1)K+k_n}+s_{mn}^{(k_m,k_n)}\notag\\
 &= h_{n,(i-1)K+k_m}^Tw_{m,(j-1)K+k_n}+s_{nm}^{(k_n,k_m)},
\end{align}
following \eqref{eq:h_vector_dedicated}. Next, taking their difference we get,
\begin{align}
    h_{n,(i-1)K+k_n}^T(w_{m,(j-1)K+k_n}-w_{n,(i-1)K+k_m})=  h_{n,(i-1)K+k_n}^T(w_{\bm{v}^*}(i_{nm,1}) - w_{\bm{v^*}}(i_{nm,2})),
\end{align} 
from which we recover $w_{\bm{v^*}}(i_{nm,2})$ (or  $w_{\bm{v^*}}(i_{nm,1})$), by canceling out $w_{\bm{v^*}}(i_{nm,1})$ (or $w_{\bm{v^*}}(i_{nm,2})$), respectively, decoded in the first step, since $ h_{n,(i-1)K+k_n}$ is available to the user. Thus far, the user has decoded all $w_{\bm{v^*}}(j)$, $j\in [2D]$. Finally, we evaluate the answers from each pair of servers in $\mathcal{C}'_{D}$, similar to the decoding approach in \cite{aliDAPAC}. For each $\{n,m\}\in \mathcal{C}'_{D}$, consider the difference of answers corresponding to $\mathcal{U}_{nm}(k_n,k_m)$ and $ \mathcal{U}_{mn}(k_m,k_n)$ from servers $n$ and $m$, where $n<m$, respectively, 
\begin{align}
     h_{m,(j-1)K+k_m}^T&w_{m,(j-1)K+k_n}+s_{mn}^{(k_m,k_n)}-h_{n,(i-1)K+k_m}^Tw_{n,(i-1)K+k_m}+s_{nm}^{(k_n,k_m)}\notag\\
     &= (h_{n,(i-1)K+k_m}+e_{l_m})^Tw_{m,(j-1)K+k_n}-h_{n,(i-1)K+k_m}^Tw_{n,(i-1)K+k_m}\label{eq:resolve h_j}\\
     &= e_{l_m}^T w_{n,(i-1)K+k_m}=w_{\bm{v^*}}(i_{nm}),
\end{align}
where \eqref{eq:resolve h_j} follows from \eqref{eq:h_vector_dedicated}. This way, the $\frac{D^2-3D}{2}$ sub-packets of $W_{\bm{v}^*}$ corresponding to $\mathcal{C}'_{D}$ are decoded by the user.

The user downloads $K(D-1)$ sub-packets from each dedicated server, and $KD$ sub-packets from the central server. In total, the scheme requires $KD^2$ sub-packets, to decode $M$ sub-packets, resulting in the rate,
\begin{align}
    R=\frac{\binom{D+1}{2}}{KD^2}=\frac{D+1}{2KD}.
\end{align}
The load ratio for the scheme is $\ell=\frac{K(D-1)}{KD}=\frac{D-1}{D}$.

\paragraph{Attribute Privacy} For each dedicated server, the user requests for $K(D-1)$ linear combinations, one corresponding to each message group. With this structure, server $n$ observes $(D-1)$ sub-packets of each accessible message being requested. Furthermore, the order of the sub-packet indices are hidden through independent permutations $\pi_{\bm{v}}$, and are independent of $\bm{v}^*$.  The same arguments follow for the central server, with $D$ sub-packets of each accessible message being queried in the form of linear combinations. Finally, since the servers are non-colluding, the corresponding linear combining vectors appear to be uniformly random vectors from 
$\mathbb{F}_q$. As a result, the query tuples reveal no information to any individual server, and \eqref{eq:user_priv1} and \eqref{eq:user_priv2} hold. 

\paragraph{Database Secrecy}
Each answer sent by the dedicated server $n$ is a linear combination of the message sub-packets, mixed with a randomness sub-packet $s_{nm}^{(k_n,k)}$, where $\{n,m\}\in \mathcal{P}_D$ and $k\in [K]$. One can get some information about a given linear combination, if multiple answers from the dedicated servers utilize the same randomness sub-packet. Consider the answers of servers $n$ and $m$ in $[D]$, corresponding to $\mathcal{U}_{nm}(k_n,k_m)$. The same randomness sub-packet, $s_{nm}^{(k_n,k_m)}=s_{mn}^{(k_m,k_n)}$ is used in their answers. After processing them, the user decodes a sub-packet of $W_{\bm{v^*}}$, whereas the remaining sub-packets in the message group are protected by $s_{nm}^{(k_n,k_m)}$ by one-time padding \cite{shannon_otp}. For any pair $(k,k')$ with $k\neq k_n$ and $k'\neq k_m$, the linear combinations do not involve a sub-packet of $W_{\bm{v}^*}$. For instance, consider the linear combination arising from the message subset $\mathcal{U}_{nm}(k_n,k')$, where $k'\neq k_m$. The user receives the linear combination, mixed with the randomness $s_{nm}^{(k_n,k')}$. On the other hand, server $m$ only learns $v^*_m$; hence neither of its message groups correspond to $\mathcal{U}_{mn}(k_m,k)$, since $k'\neq k_m$. As a result, the user does not download a second linear combination, mixed with $s_{nm}^{(k_n,k')}$ if $k'\neq k_m$. This prevents the user from learning any information on the accessible messages from the answers received from the dedicated servers.

In our scheme, the central server combines the $K$ randomness sub-packets according to \eqref{eq:randomness_sum}. Consider the linear combinations corresponding to $\mathcal{U}(n,k_n)$, that are mixed with the randomness sub-packet $\tilde{s}_{D+1,(n-1)K+k_n}$. By adding $K$ answers of the dedicated server $n$ according to \eqref{eq:sum_dedicated}, one can decode a sub-packet of the designated message. However, since $s_{nm}(k_n,k)$ for all  $(n,m)\in \mathcal{C}^o_{D}$ and $k\in [K]$, are unavailable to the user, the information in the remaining message sub-packets in the $K$ message groups of server $n$ are perfectly protected from the user \cite{shannon_otp}. Finally, for the linear combinations corresponding to $\mathcal{U}(n,k)$ where $k\neq k_n$, we write for $\{n,m\}\in \mathcal{C}_{D}$,
\begin{align}
    \mathcal{U}(n,k)=\bigcup_{k'=1}^K \mathcal{U}_{mn}(k',k)= \mathcal{U}_{mn}(k_m,k)\cup\left(\bigcup_{k'=1, k;\neq k_m}^K \mathcal{U}_{mn}(k',k)\right).
\end{align}
Since server $m$ verifies $v^*_m$, the linear combinations corresponding to $\mathcal{U}_{mn}(k',k), k\neq k_m$ are not requested by the user. As a result, the user does not download answers using the randomness sub-packets  $s_{mn}^{(k',k)}, k\neq k_m$. This safeguards the secrecy of the message sub-packets involved in these answers of the central server. 

\subsection{Example~\ref{ex:3} Revisited}
For ease of understanding, we revisit Example~\ref{ex:3} in light of the notations introduced in the scheme. We have $\mathcal{P}_3=\{\{1,2\},\{1,3\},\{2,3\}\}=\mathcal{C}_{3}$ and $\mathcal{C}'_{3}=\emptyset$. Also, let $\mathcal{C}_{3}^o=\{(1,2), (2,3), (3,1)\}$. Then, the message sets corresponding to $\{1,2\}\in \mathcal{P}_3$ are,
    \begin{align}
    &\mathcal{U}_{12}(1,1)= \mathcal{U}_{21}(1,1) = \{W_{a1uy},W_{a1vy}\},\notag\\
    &\mathcal{U}_{12}(1,2)= \mathcal{U}_{21}(2,1) = \{W_{a2uy},W_{a2vy}\},\notag\\
    &\mathcal{U}_{12}(2,1)= \mathcal{U}_{21}(1,2) = \{W_{b1uy},W_{b1vy}\},\notag\\
    &\mathcal{U}_{12}(2,2)= \mathcal{U}_{21}(2,2) = \{W_{b2uy},W_{b2vy}\}.
    \end{align}
The message sets corresponding to $\{1,3\}\in \mathcal{P}_3$ are,
    \begin{align}
    &\mathcal{U}_{13}(1,1)= \mathcal{U}_{31}(1,1) = \{W_{a1uy},W_{a2uy}\},\notag\\
    &\mathcal{U}_{13}(1,2)= \mathcal{U}_{31}(2,1) = \{W_{a1vy},W_{a2vy}\},\notag\\
    &\mathcal{U}_{13}(2,1)= \mathcal{U}_{31}(1,2) = \{W_{b1uy},W_{b2uy}\},\notag\\
    &\mathcal{U}_{13}(2,2)= \mathcal{U}_{31}(2,2) = \{W_{b1vy},W_{b2vy}\}.
    \end{align}
The message sets corresponding to $\{2,3\}\in \mathcal{P}_3$ are,
    \begin{align}
    &\mathcal{U}_{23}(1,1)= \mathcal{U}_{32}(1,1) = \{W_{a1uy},W_{b1uy}\},\notag\\
    &\mathcal{U}_{23}(1,2)= \mathcal{U}_{32}(2,1) = \{W_{a1vy},W_{b1vy}\},\notag\\
    &\mathcal{U}_{23}(2,1)= \mathcal{U}_{32}(1,2) = \{W_{a2uy},W_{b2uy}\},\notag\\
    &\mathcal{U}_{23}(2,2)= \mathcal{U}_{32}(2,2) = \{W_{a2vy},W_{b2vy}\}.
    \end{align}
Each message set corresponds to a randomness sub-packet from $\mathcal{S}$, which are relabeled in the example as follows,
\begin{align}
    &s_{12}^{(1,1)}=s_{21}^{(1,1)}=s_1,  
    &&s_{12}^{(1,2)}=s_{21}^{(2,1)}=s_2, \notag\\
    &s_{12}^{(2,1)}=s_{21}^{(1,2)}=s_{10}, 
    &&s_{12}^{(2,2)}=s_{21}^{(2,2)}=s_5, \notag\\
    &s_{13}^{(1,1)}=s_{31}^{(1,1)}=s_3,  
    &&s_{13}^{(1,2)}=s_{31}^{(2,1)}=s_4, \notag\\
    &s_{13}^{(2,1)}=s_{31}^{(1,2)}=s_9, 
    &&s_{13}^{(2,2)}=s_{31}^{(2,2)}=s_{12}, \notag\\
    &s_{23}^{(1,1)}=s_{32}^{(1,1)}=s_8, 
    &&s_{23}^{(1,2)}=s_{32}^{(2,1)}=s_{11}, \notag\\
    &s_{23}^{(2,1)}=s_{32}^{(1,2)}=s_6,  
    &&s_{23}^{(1,2)}=s_{32}^{(2,2)}=s_7,.
\end{align}
Towards query construction, we have $v_1^*=a=\mathcal{V}_1(1)$, i.e., $k_1=1$, $v_2^*=2=\mathcal{V}_2(2)$, i.e., $k_2=2$ and $v^*_3=\mathcal{V}_3(1)$, i.e., $k_3=1$. Accordingly, the message groups assigned for server $1$ in Table \ref{tab:ex2_3} correspond to $\mathcal{U}_{12}(1,1)$, $\mathcal{U}_{12}(1,2)$, $\mathcal{U}_{13}(1,1)$ and $\mathcal{U}_{13}(1,2)$, with one new sub-packet index assigned per message. Next, the message groups assigned for server $2$ in Table \ref{tab:ex2_3} correspond to $\mathcal{U}_{21}(1,1)$, $\mathcal{U}_{21}(2,1)$, $\mathcal{U}_{23}(1,1)$ and $\mathcal{U}_{23}(2,1)$, while those for server $3$ are  $\mathcal{U}_{31}(1,1)$, $\mathcal{U}_{31}(2,1)$, $\mathcal{U}_{32}(1,1)$ and $\mathcal{U}_{32}(2,1)$. The sub-packet indices assigned for $W_{\bm{v^*}}$, corresponding to the elements in $\mathcal{C}_{3}$ are $i_{12,1}=1$, $i_{12,2}=4$, $i_{13,1}=2$, $i_{13,2}=5$, $i_{23,1}=3$, $i_{23,2}=6$. The message groups of the dedicated servers are labeled $w_{n,l}$, $n\in [3]$, $l\in[4]$ following \eqref{eq:label_msg_grp} with $\bar{1}=(2,3)$, $\bar{2}=(1,3)$ and $\bar{3}=(1,2)$. The corresponding linear combining vectors are assigned by uniformly chosen vectors from $\mathbb{F}_q^{2}$ and chosen independently, for all $h_{n,l}$ except $h_{2,1}=h_{1,2}$, $h_{3,1}=h_{1,3}$ and $h_{3,4}=h_{2,3}$, following \eqref{eq:h_vector_dedicated}. For the central server, the message groups corresponding to $\mathcal{U}(1,k_1)=\mathcal{U}(1,1)$, $\mathcal{U}(2,k_2)=\mathcal{U}(2,2)$, and $\mathcal{U}(3,k_3)=\mathcal{U}(3,1)$, i.e., $w_{4,1}$, $w_{4,4}$, $w_{4,5}$ are formed according to \eqref{eq:msg_set_concat}. The remaining message groups $w_{4,2}, w_{4,3}, w_{4,6}$ are formed by picking new sub-packets of messages in $\mathcal{U}(1,2)$, $\mathcal{U}(2,1)$ and $\mathcal{U}(3,2)$, respectively. The linear combining vectors $h_{4,2}$, $h_{4,3}$, $h_{4,6}$ are assigned uniformly at random from the $\mathbb{F}_q^4$, while the remaining vectors are set according to \eqref{eq:h_vector for D+1} as follows,
\begin{align}
   h_{4,1}=[h_{1,1};h_{1,2}+e_1], \quad h_{4,4}=[h_{2,3}+e_1;h_{2,4}], \quad h_{4,5}= [h_{1,3}+e_2;h_{3,2}],
\end{align}
since $h_{1,3}=h_{3,1}$. This completes the generation of query tuples. 

The answer generation is straightforward for the dedicated servers. For the central server, sums of randomness sub-packets are formed following the elements of $\mathcal{C}^o_{3}$. For $(1,2)$, we have
\begin{align}
    &\tilde{s}_{4,1}=s_{12}^{(1,1)}+s_{12}^{(1,2)}= s_1+s_2, \\
    &\tilde{s}_{4,2}=s_{12}^{(2,1)}+s_{12}^{(2,2)}= s_{10}+s_{5}, 
\end{align}
for $(2,3)$, we have
\begin{align}
    &\tilde{s}_{4,3}=s_{23}^{(1,1)}+s_{23}^{(1,2)}= s_8+s_{11}, \\
    &\tilde{s}_{4,4}=s_{23}^{(2,1)}+s_{23}^{(2,2)}= s_{6}+s_{7}, 
\end{align}
and for $(3,1)$, we have
\begin{align}
    &\tilde{s}_{4,5}=s_{31}^{(1,1)}+s_{31}^{(1,2)}= s_3+s_{9}, \\
    &\tilde{s}_{4,6}=s_{31}^{(2,1)}+s_{31}^{(2,2)}= s_{4}+s_{12}.   
\end{align}
This completes answer generation for the central server.

\subsection{Proof of Remark~\ref{rem4}}
Using \eqref{eq:tradeoff} with $\ell(\lambda)= \frac{D-1}{D}$, we obtain the rate through time-sharing to be
\begin{align}
   \frac{KD+K-1}{2K^2D} = \frac{D+1}{2KD} - \frac{1}{2K^2D} <\frac{D+1}{2KD}.
\end{align}
Further, the gain in rate is $\frac{1}{2K^2D}$, which diminishes with increasing $K$ and $D$. 

\section{Conclusion}\label{sec:conclude}
In this work, we proposed an attribute verification and message retrieval framework, by augmenting a central server to the existing $(D,K)$ DAPAC system which was proposed to verify $D$ private attributes. While the $(N,D,K)$ system model requires the knowledge of $N-D$ attributes by the $D+1$ authorities post-verification, we show that it improves the achievable rate by a factor of $\frac{2K}{K+1}$, i.e., almost by a factor of two. We studied the asymmetric load ratio resulting from our model, and established a rate-load ratio trade-off through time-sharing. Further, we proposed an improved achievable $(N,D,K)$ scheme with the load ratio fixed to $\frac{D-1}{D}$, resulting in an improved rate-load ratio trade-off.

The capacity of the DAPAC problem is a trivial upper bound on the capacity of the HetDAPAC problem, because any $(D,K)$ DAPAC scheme is a feasible $(N,D,K)$ HetDAPAC scheme, where the central server does not participate in the message retrieval phase. Beyond this, the question of optimality of these schemes has not been explored, and establishing upper bounds on the capacity of DAPAC and HetDAPAC problems is an interesting direction for future research. Similarly, the optimality of the rate-load ratio trade-off is open for future work. Moreover, the assumption that the central server relays all the public attributes after verification to the remaining servers, is crucial to obtaining gain in the achievable rates. As a result, each attribute has two privacy levels, 1) be revealed to a single server, or 2) be revealed to all servers. The scenario of multiple privacy levels, where the privacy level of a given attribute is determined by the number of servers that it can be revealed to, presents another generalization of this work.  

\bibliographystyle{unsrt}
\bibliography{reference}
\end{document}